\documentclass[12pt]{article}
\pdfoutput=1
\usepackage{graphicx}
\usepackage{dcolumn}

\newcommand\pubnumber{DPF2013-72}
\newcommand\pubdate{\today}

\def\UVa{Department of Physics\\
University of Virginia, Charlottesville, VA, 22904, USA}

\def\Title#1{\begin{center} {\Large #1 } \end{center}}
\def\Author#1{\begin{center}{ \sc #1} \end{center}}
\def\Address#1{\begin{center}{ \it #1} \end{center}}

\newcommand\pubblock{\rightline{\begin{tabular}{l} \pubnumber\\
         \pubdate  \end{tabular}}}
\newenvironment{Abstract}{\begin{quotation}  }{\end{quotation}}
\newenvironment{Presented}{\begin{quotation} \begin{center} 
             PRESENTED AT\end{center}\bigskip 
      \begin{center}\begin{large}}{\end{large}\end{center} \end{quotation}}
\def\Acknowledgments{\bigskip  \bigskip \begin{center} \begin{large}
             \bf ACKNOWLEDGMENTS \end{large}\end{center}}

\def\beq{\begin{equation}}
\def\eeq#1{\label{#1}\end{equation}}
\def\eeqn{\end{equation}}

\def\beqa{\begin{eqnarray}}
\def\eeqa#1{\label{#1}\end{eqnarray}}
\def\eeqan{\end{eqnarray}}

\let\bar=\overbar

\def\Dslash{\not{\hbox{\kern-4pt $D$}}}
\def\dslash{\not{\hbox{\kern-2pt $\del$}}}

\def\msb{{\bar{\ssstyle M \kern -1pt S}}}

\begin{document}
\newcommand{\met}{\ensuremath{\not \!\! E_T}}
\newcommand{\MET}{\ensuremath{\not \!\! E_T}}
\newcommand{\isotrk}{IsoTrk}

\newcommand{\METraw}{E_{T}^{\rm{raw}}\!\!\!\!\!\!\!\!\!\!\! / \;\;\;\;}
\newcommand{\vMET}{\vec{E}_{T}\!\!\!\!\!\! /\;\;}
\newcommand{\invfb}{\mbox{fb}^{-1}}
\newcommand{\pt}{p_{T}}
\newcommand{\et}{E_{T}}
\newcommand{\tbar}{\bar{\text{t}}}

\newcommand{\GeV}{~\mbox{GeV}}
\newcommand{\fb}{~\mbox{fb}}
\newcommand{\secvtx}{{\sc Secvtx}}
\newcommand{\pythia}{{\sc Pythia}}
\newcommand{\powheg}{{\sc Powheg}}
\newcommand{\alpgen}{{\sc Alpgen}}
\newcommand{\metsig}{MET_{\mbox{sig}}}
\newcommand{\branchingratio}{\mbox{Br}}

\begin{titlepage}
\pubblock

\vfill
\Title{Evidence of s-channel Single Top Quark Production in Events with one Charged Lepton and Bottom Quark Jets at CDF}
\vfill
\Author{Hao Liu\\
(On Behalf of the CDF Collaboration)}
\Address{\UVa}
\vfill
\begin{Abstract}
We report an evidence of \textit{s}-channel single top quark production in $p\bar{p}$ collision at $\sqrt{s}= 1.96~\mathrm{TeV}$ using data with integrated luminosity of $9.4~\mathrm{fb}^{-1}$ collected by the Collider Detector at Fermilab (CDF\cite{CDF}). We select events with one charged lepton, large missing transverse energy and two bottom quark jets. The observed significance of the result is $3.8$ standard deviation from background only prediction. We measure the inclusive cross section to be $1.41^{+0.44}_{-0.42}\mathrm{(stat+syst)}~\mathrm{pb}$ assuming $m_t = 172.5~\mathrm{GeV}/c^2$.
\end{Abstract}
\vfill
\begin{Presented}
DPF 2013\\
The Meeting of the American Physical Society\\
Division of Particles and Fields\\
Santa Cruz, California, August 13--17, 2013\\
\end{Presented}
\vfill
\end{titlepage}
\def\thefootnote{\fnsymbol{footnote}}
\setcounter{footnote}{0}

\section{Introduction}

\begin{figure}[htbp]
  \begin{center}
    \includegraphics[width=0.3\textwidth]{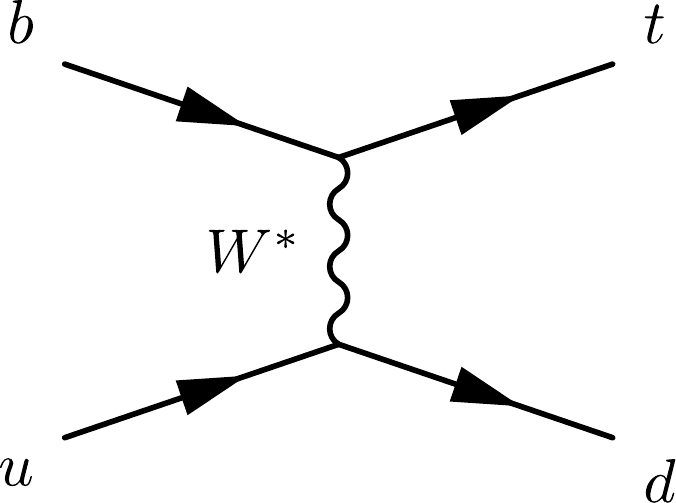}
    \includegraphics[width=0.3\textwidth]{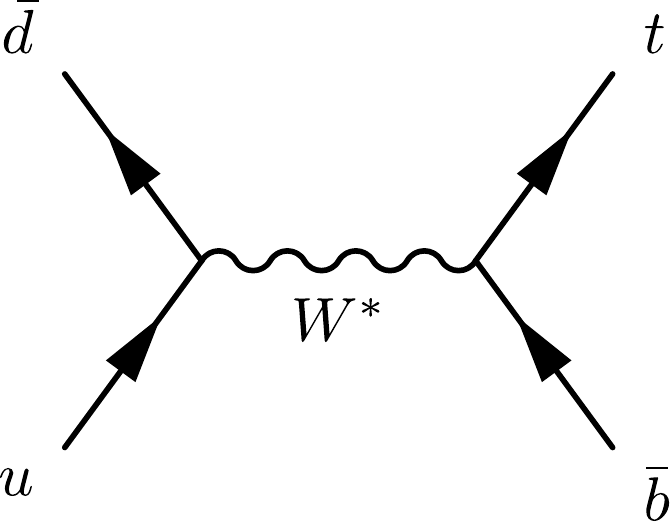}
    \includegraphics[width=0.3\textwidth]{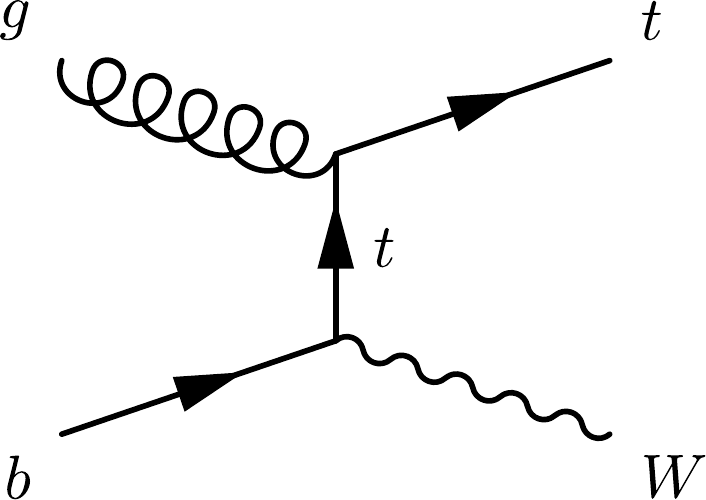}
    \caption{Feynman diagram for the single top process. The left one is $t$-channel process, the center one is $s$-channel process and the right one is $tW$-channel process.}
    \label{fig:STFeynmanDiagram}
  \end{center}
\end{figure}

In the Standard Model(SM), the top quark can be produced not only in pair by strong interactions but also singly through weak interactions. At hadron collider, there are three different modes for single top production. The first one is an intermediate $W$ boson decays into a top(antitop) quark and a antibottom(bottom) quark($s$-channel). The second one is a bottom quark transforms into a top quark by exchanging a $W$ boson with another quark($t$-channel). The last one is a top quark produced associate with a $W$ boson($tW$-channel). As shown in Figure~\ref{fig:STFeynmanDiagram}. Because of the short life time and heavy mass of top quarks, the single top process provides a unique opportunity to test the SM and search for new physics.

At Tevatron, the cross section for $s$-channel process is so small that only recently D0 claimed an evidence of this process~\cite{D0}. Moreover, Tevatron experiments are more sensitive to this process than LHC experiments since LHC is a $pp$ collider.

\section{Data Sample and Event Selection}
\label{sec:event}

In this analysis, we select events consistent with a $W$-boson decays into a charged lepton and corresponding neutrino plus two energetic $b$-quark jets. Since this process shares the same final state as $WH$ process at Tevatron, in this analysis, we use the same event selection as $WH$ analysis~\cite{WH}.

We require one single, isolated lepton with $\pt > 20\ \GeV/c$, and the presence of $\MET$ in the event. We use different lepton reconstruction algorithms for leptons collected by different parts of our detector, and we divide them into following categories to keep them orthogonal to each other, as list below.

\begin{itemize}
\item CEM: central tight electron
\item CMUP and CMX: central tight muon
\item Extended Muon Category(EMC): loose muons and reconstructed isolated track lepton candidates
\end{itemize}

We also apply different $\met$ thresholds depending on the charged lepton category.
We require  $\met > 10 \GeV$ for CMUP and CMX events, $\met > 20 \GeV$ for CEM and EMC events.

Jets information used in the analysis are reconstructed with the JetClu algorithm with a cone size of 0.4. Selected jets are required to have corrected $\et > 20 \GeV/c^2$ and $\eta < 2.0$. Only events with exactly two  jets are accepted. 
We also employ a $b$-tagging algorithm to further select our events. The $b$-tagging algorithm is denoted as The HOBIT~\cite{HOBIT}. We require at least one of the jets to be tagged by HOBIT. We defined two operational points of the tagging algorithm based on the output value of HOBIT, Tight, output larger than 0.98 and Loose, output larger than 0.72.
Based on the tagging information, we divide our events into following four orthogonal tagging categories:

\begin{itemize}
\item TT: Exactly two jets tagged by HOBIT Tight
\item TL: One jet is HOBIT Tight, another jet is HOBIT Loose, but not HOBIT Tight
\item T: One of the jets is HOBIT Tight, other jets are not HOBIT Loose
\item LL: None of the jets are HOBIT Tight, exactly two jets are HOBIT Loose
\end{itemize}

\section{Backgrounds Estimation}
\label{sec:background}
We determine the fraction of $W$ + jets events for each lepton category in pretag regionby fitting the $\met$ distribution of pretag samples. For single top, $t\bar{t}$, diboson and $Z$ + jets samples are normalized to their corresponding theoretical expectation, while $W$ + jets and multijet QCD samples normalization are free to float in the likelihood fit.

The normalization of $W$ + jets sample in tagged region are calculated from the pretag region by applying heavy flavor fraction and tagging efficiency ($W$ + heavy flavor) or light flavor fraction and mistag matrix($W$ + light flavor). Multijet samples are calculated by applying tag rate calculated from data.

The prediction for number of events in each tagging category are shown in Table~\ref{tab:background}.

\begin{table}[htbp]
  \begin{center}
    \newcolumntype{d}{D{,}{\pm}{-1}}
    \begin{tabular}{cdddd}
      \hline
      Category            & \multicolumn{1}{c}{TT}  & \multicolumn{1}{c}{TL} & \multicolumn{1}{c}{T} & \multicolumn{1}{c}{LL} \\  
      \hline           
      $WW$                & 1.7 , 0.4     &  13.2 , 2.7   &   184 , 23        & 24.8 , 3.9    \\  
      $WZ$                & 17.8 , 2.2    &  21.2 , 2.0   &   52.7 , 5.4      & 9.9 , 0.9     \\  
      $ZZ$                & 2.4 , 0.3     &  2.4 , 0.2    &   7.1 , 0.7       & 0.96 , 0.08     \\  
      $Z$ + jets          & 10.9 , 1.2    &  20.7 , 2.3   &   163 , 18        & 27.1 , 3.1    \\  
      $t\bar{t}$          & 163 , 21      &  194 , 19     &   502 , 50        & 58.1 , 6.6      \\  
      Higgs               & 6.1 , 0.6     &  6.4 , 0.4    &   10.3 , 0.7      & 1.7 , 0.2     \\  
      $Wbb$               & 246 , 99      &  327 , 130    &   1166 , 468      & 109 , 44   \\  
      $Wcc$               & 19.0 , 7.8    &  120 , 49     &   1158 , 467      & 164 , 67   \\  
      $W$ + Mistag        & 4.3 , 1.3     &  62 , 13      &   978 , 141       & 242 , 34   \\  
      Multijet            & 29 , 12       &  47 , 19      &   281 , 112       & 45 , 18    \\  
      $t$ and $Wt$-channel & 18.1 , 2.5  &  35.3 , 4.2   &   251 , 28        & 13.6 , 1.5    \\  
      \hline 
      $s$-channel         & 54.5 , 6.7    &  61.2 , 5.6   &   109 , 10        & 17.8 , 2.1    \\  
      \hline 
      Total Prediction    & 573 , 155     &  911 , 248    &   4860 , 1320     & 714 , 181    \\  
      \hline 
      Observed            & \multicolumn{1}{c}{466} &  \multicolumn{1}{c}{765} & \multicolumn{1}{c}{4620} & \multicolumn{1}{c}{718}  \\  
      \hline
    \end{tabular}
    \caption{Summary of background and signal prediction in each tagging category, with systemactic uncertainties of cross section included.}
    \label{tab:background}
  \end{center}
\end{table}

\section{Final Discriminant}
\label{sec:discr}
\begin{figure}[htbp]
  \begin{center}
    \includegraphics[width=0.45\textwidth]{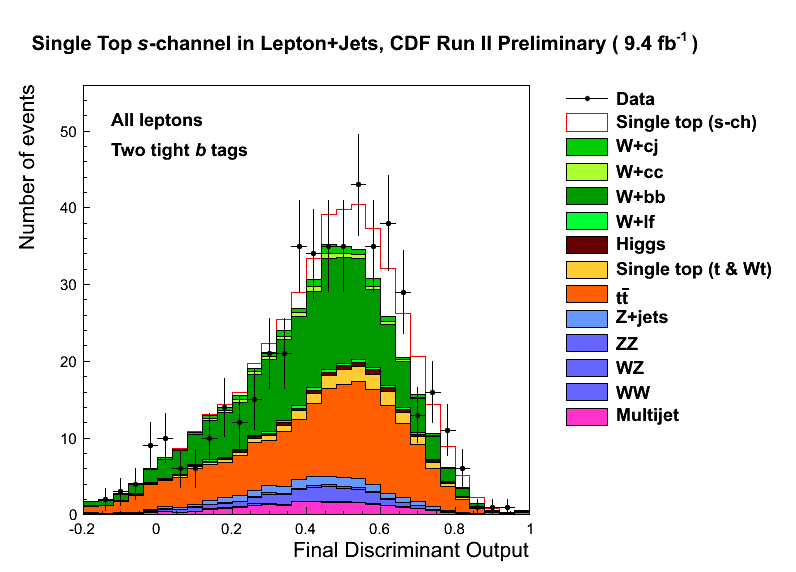}
    \includegraphics[width=0.45\textwidth]{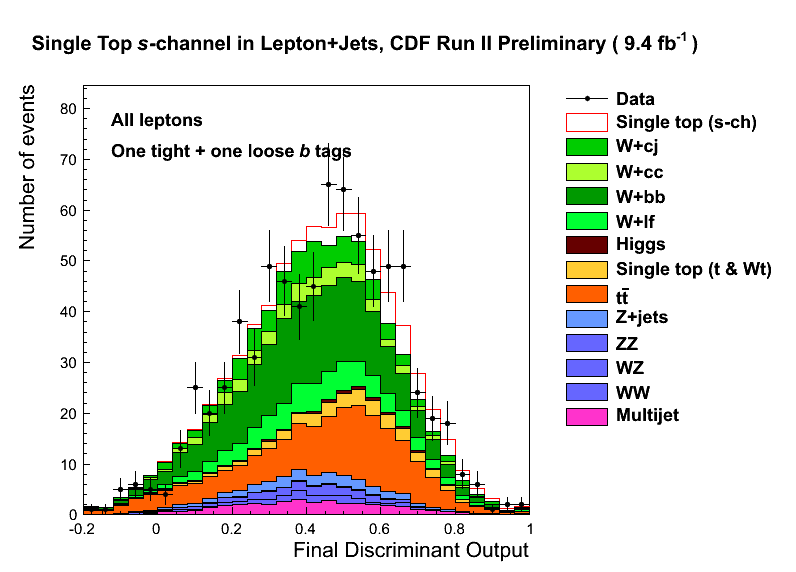}
    \caption{Final discriminant output for TT and TL tagging category and all lepton category combined}
    \label{fig:discriminant1}
  \end{center}
\end{figure}

To further separate the signal from background, and increase the sensitivity of this analysis, we use TMVA package trained a neural network to be the final discriminant.We trained separate neural networks for each tagging category. 
The final discriminant output distributions of TT and TL tagging category are shown in Figure~\ref{fig:discriminant1}. The description of input variables used in the final discriminant are listed below.

\begin{description}
\item[$M_{l\nu b}$]  The reconstructed top quark mass
\item[$M_{l\nu bb}$] The reconstructed mass of the charged lepton, $\met$ and two jets
\item[Lep $\pt$] The $\pt$ of the charged lepton
\item[$M_{jj}$] The reconstructed mass of two jets corrected using neural network~\cite{NIMbjetcorr}
\item[$\mathrm{cos}\theta_{lj}$] The cosine of the angle between the charged lepton and the jet selected to reconstruct top quark in the top quark rest frame
\item[$H_t$] The scalar sum of transverse energy of the charged lepton, $\met$ and all jets
\item[$M_{l\nu b}^T$] The transverse mass of the reconstructed top quark
\item[$b$ jet selector output] The output of the neural network used to select the $b$ jet originated from top quark
\end{description}

\section{Measurement}
\label{sec:measurement}
We measure the single top cross section using a Bayesian binned
likelihood technique~\cite{PDG} assuming a flat prior in the cross section
and integrating the posterior over all sources of systematic
uncertainty.

\begin{figure}[htbp]
  \begin{center}
    \includegraphics[width=0.45\textwidth]{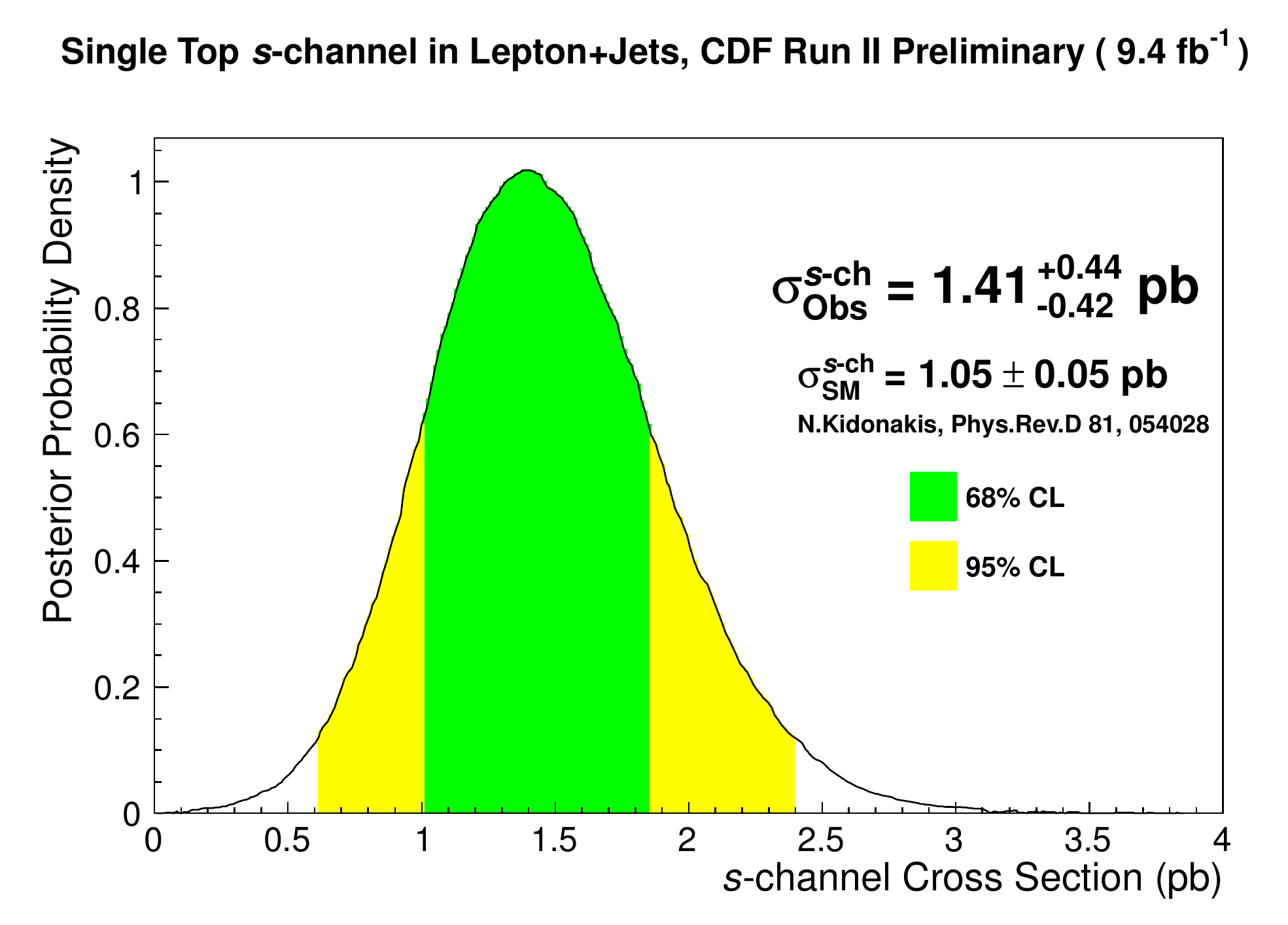}
    \includegraphics[width=0.45\textwidth]{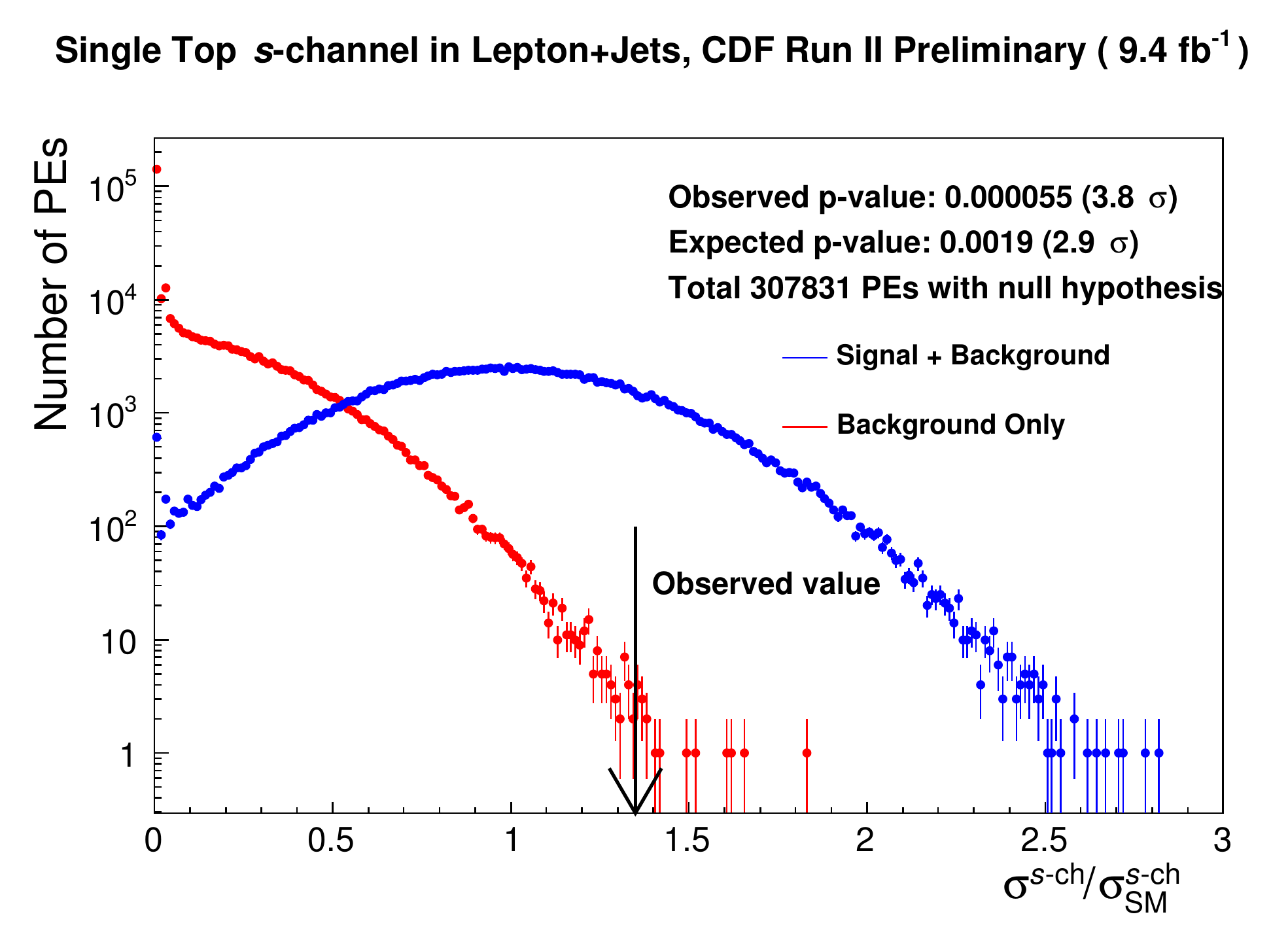}	
    \caption{The left figure shows the posterior probability density distribution for the cross section measurement. The green and yellow regions represent the smallest intervals enclosing 68\% and 95\% of the posterior density integrals. The right figure shows pseudo-experiment output with background only hypothesis and background plus signal hypothesis. The arrow is pointing at the measured cross section value.}
    \label{fig:xsec}
  \end{center}
\end{figure}

The posterior probability distribution of single top $s$-channel cross section from calculation is shown in Figure~\ref{fig:xsec}. From the distribution, we measure the $s$-channel cross section to be $\sigma_{s\mathrm{-channel}}=1.41^{+0.44}_{-0.42}~\mathrm{pb}$.

We also measured the p-value by generating pseudo-experiment with both background only and signal plus background hypothesis, as shown in Figure~\ref{fig:xsec}. The p-value for the observed cross section is 0.0000597, which corresponds to a significance of 3.8$\sigma$.

\section{Conclusion}
\label{sec:conclusion}

We have presented the results of a search for the single top $s$-channel. We find that for the dataset corresponding to integrated luminosity of 9.4~$\invfb$, the data agrees with the Standard Model background predictions within the systematic uncertainties. 

We measure the single top $s$-channel cross section to be $\sigma_{s\mathrm{-channel}}=1.41^{+0.44}_{-0.42}~\mathrm{pb}$, assuming the top quark mass is $172.5~\mathrm{GeV}/c^2$. This corresponds to a significance of 3.8$\sigma$. The results is compatible with standard model prediction and is also compatible with previous CDF measurements.

\Acknowledgments
We thank the Fermilab staff and the technical staffs of the
participating institutions, and funding agencies for their vital contributions.

\end{document}